\documentclass{svjour2}                    
\smartqed  
\usepackage{graphicx}
\usepackage{natbib}
\usepackage{amssymb,amsmath,graphicx}
\usepackage{units}
\usepackage{upgreek}
\usepackage{subfig}

\newcommand{\bee}{\begin{equation}}
\newcommand{\ee}{\end{equation}}

\newcommand{\rd}{\ensuremath{\mathrm{d}}}

\renewcommand{\vec}[1]{\mathbf{#1}}

\journalname{Acta Mechanica}
\begin{document}
\title{Periodic and aperiodic tumbling of microrods advected in a microchannel flow} 
\author{J. Einarsson \and A. Johansson \and  S.K. Mahato \and Y.N. Mishra \and J.R. Angilella \and D. Hanstorp \and B. Mehlig}
\institute{J. Einarsson \and A. Johansson \and  S.K. Mahato \and Y.N. Mishra \and D. Hanstorp \and B. Mehlig
                \at Department of Physics, University of Gothenburg, 412 96 Gothenburg, Sweden
           \and
           J. R. Angilella \at LUSAC, Universit\'e{} de Caen, Cherbourg, France
}
\date{Received: date / Accepted: date}
\maketitle
\begin{abstract}
We report on an experimental investigation of
the tumbling of microrods in the shear flow of a microchannel
(dimensions: $\unit[40]{mm} \times \unit[2.5]{mm} \times \unit[0.4]{mm}$).
The rods are $\unit[20]{\upmu m}$ to $\unit[30]{\upmu m}$ long and their diameters are of
the order of $\unit[1]{\upmu m}$. Images of the centre-of-mass motion and the orientational dynamics of the rods are recorded using a microscope
equipped with a CCD camera. A motorised microscope stage is used to track individual rods as they move along the channel.
Automated image analysis determines the position and orientation of a tracked rods in each video frame.
We find different behaviours, depending on the particle shape, 
its initial position, and orientation. First, we observe periodic as well as aperiodic tumbling. Second, the data show that different tumbling trajectories exhibit
different sensitivities to external perturbations.  These observations can be explained by slight asymmetries of the rods.
Third we observe that after some time, initially periodic trajectories lose their phase.
We attribute this to drift of the centre of mass of the rod from one to another stream line of the channel flow.
\end{abstract}

\maketitle 
\section{Introduction}\label{sec:introduction}
It was shown by \citet{jeffery1922} that small axisymmetric rods in a viscous shear flow align for the most time with 
the flow direction, but that their symmetry axes periodically and rapidly turn by 180 degrees. 
This motion is referred to as \lq tumbling\rq{} in the literature. Experimental studies of tumbling of particles in flows 
have been performed for a long time, see for example \citet{goldsmith1962}. More recently, \citet{kaya2009} report on
an experiment analysing the tumbling of {\em E. coli} cells in microchannel shear flows. 
They characterised the tumbling by fitting periodic orientational trajectories to short experimental time series.
A similar approach was adopted by \citet{mishra2012} to describe the tumbling of microrods in a microchannel flow. 
The periodic solutions referred to above were first obtained by \citet{jeffery1922} and are commonly referred
to as \lq Jeffery orbits\rq.

\citet{hinch1979} have shown by theoretical analysis that slightly asymmetric rods also tumble, but that their orientational 
motion is in general not strictly periodic. Under certain circumstances the authors predict that the orientational dynamics may be \lq doubly periodic\rq. 
\citet{yarin1997} refer to this motion as \lq quasi periodic\rq. Their numerical results show that the tumbling can be
chaotic \citep{Ott}. In order to experimentally distinguish periodic from aperiodic tumbling due to asymmetry, it is necessary to observe
long sequences of flips. A confounding factor is rotational and centre-of-mass diffusion: the centre of mass of an advected rod
may diffuse to neighbouring stream lines of the flow. This causes aperiodicity. 
But noise can also directly affect the rotational degrees of freedom, leading to random tumbling.

In order to disentangle these effects it is necessary to follow the orientational dynamics for many flips. The rods
must therefore be tracked for long distances. To achieve this we have improved an existing experimental setup \citep{mishra2012} in two ways. 
First, we have automated the image analysis of the empirical data. Second, with the new setup it is possible
to periodically revert the channel flow. This allows us to record longer tumbling sequences. 
In an ideal experiment the orientational dynamics
is expected to retrace its trajectory upon reversal of the flow. This is a consequence of the time-reversal invariance of the Stokes equation governing 
low Reynolds-number flow.
Our setup thus enables us to quantify the sensitivity of the observed tumbling motion to perturbations.
In the following we report and discuss experimentally observed tumbling trajectories. 
First, we see periodic and aperiodic tumbling. Second, different tumbling trajectories exhibit
different sensitivities to external perturbations. We argue that these observations can be explained by slight asymmetries of the rods.
Third, at very long times initially periodic trajectories lose their phase.
We attribute this to drift of the centre of mass of the rod from one to another stream line of the channel flow.

We conclude this introduction by briefly commenting on the relevance of the questions addressed in this paper.
Jeffery's periodic tumbling solutions in shear flows form, together with orientational diffusion, the basis of many studies of the rheology of suspensions of
non-spherical particles, as outlined by \citet{hinch1972}. \citet{petrie1999} has given an overview of the rheology of fibre suspensions and the use of
Jeffery's and diffusion theory in this field.  Furthermore, pattern formation by rods in random flows
was investigated using Jeffery's theory by \citet{wilkinson2009} and \citet{bezuglyy2010} (see also \citet{wilkinson2011}), identifying singularities in
the orientational patterns of rheoscopic suspensions, and explaining how Jeffery's periodic solutions determine rheoscopic visualisations of flows.
Last but not least, the tumbling and alignment of small rods in turbulent flows has recently been intensively
investigated, both experimentally, by simulations, and theoretically. We refer to the articles
in this special issue of Acta Mechanica, as well as 
to \citet{Parsa} and \citet{Wilkinson12}, and to articles cited in these two papers. 
Simulations and the theoretical treatments of the tumbling of small rods in turbulent
flows are based on Jeffery's equation of motion.

\section{Materials \& methods}\label{sec:method}

\subsection{Experimental methods}\label{sec:experimental_methods}
{\em Particle synthesis}.  As in our earlier experiments \citep{mishra2012}, the polymer microrods are prepared by a liquid-liquid dispersion technique using the protocol of \citet{alargova2004}.
This method produces rods with lengths of $\unit[10]{\upmu m}$ to  $\unit[100]{\upmu m}$, with typical aspect ratios of the order of 10:1. 
The resolution of the microscope does not allow us to determine whether the rods are symmetric or not, that is, whether
they have perfectly circular or slightly elliptical cross sections.
Inspection under an optical microscope shows that the shorter rods are straight. Some of the longer rods are 
curved.  In the experiments described below, rods in the shortest range of the span given above are used.

\begin{figure}
\includegraphics[width=\textwidth]{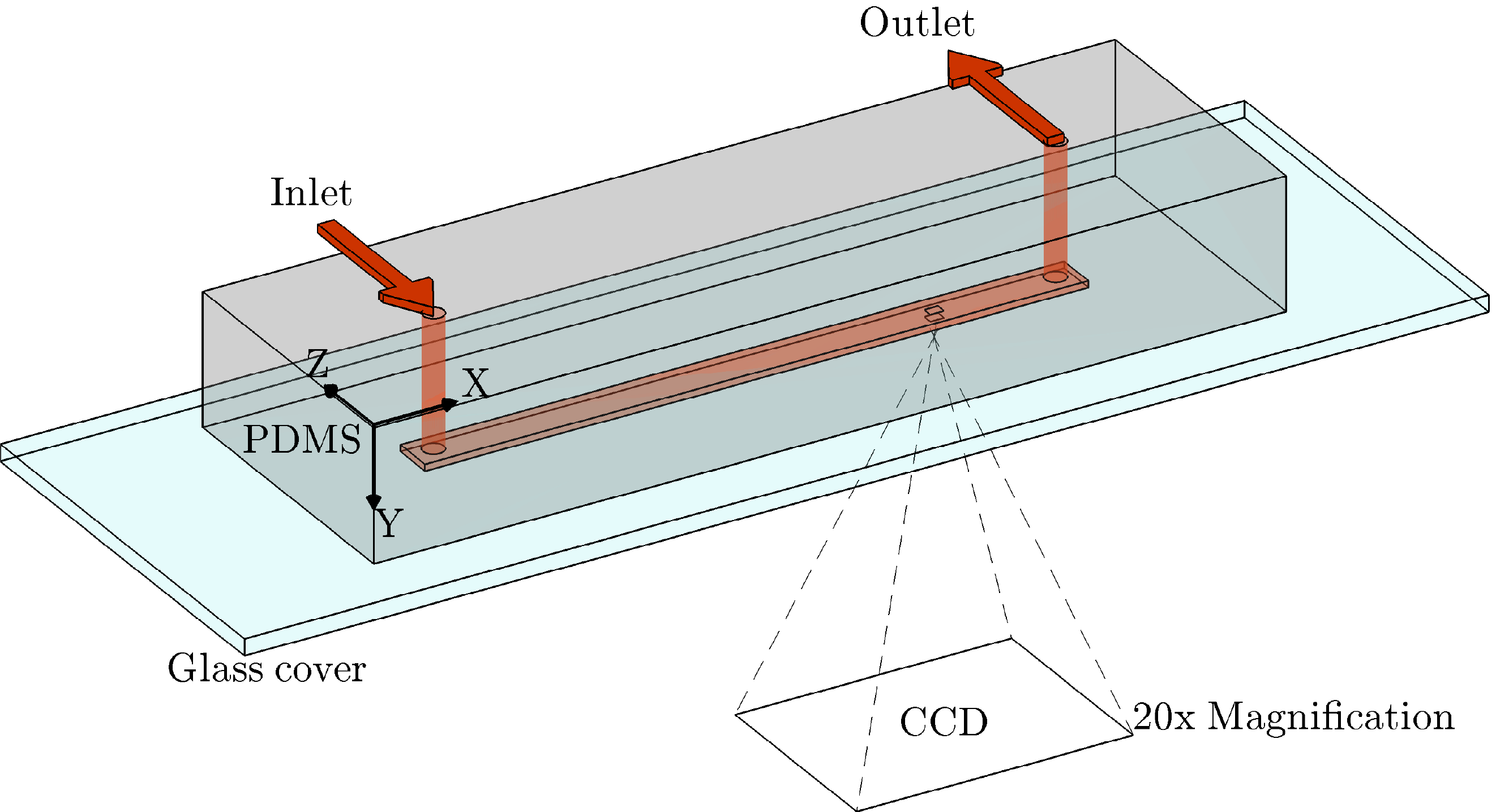}
\caption{\label{fig:experimental_setup}Experimental setup (schematic) with coordinate system. The microchannel is placed upon a motorised 
stage which allows the camera to study any part of the channel. The calculated flow rates in a cross-section of the channel are shown in Fig.~\ref{fig:flowprofile_theory}. The coordinate system used in the analysis is shown in the figure. The origin of  the coordinate system is arbitrary since only relative distances are used in the analysis.} 
\end{figure}

{\em Design and fabrication of microfluidic channels}.
The microfluidic device used in this experiment is shown schematically in Fig.~\ref{fig:experimental_setup}.
The channel ($\unit[40]{mm}$ long, $\unit[2.5]{mm}$ wide, and $\unit[0.4]{mm}$ deep) is produced in PDMS.
The procedure is described in detail by \citet{mishra2012}.

{\em Optical system and tracking}.
A sketch of the experimental setup is shown in Fig.~\ref{fig:experimental_setup}. 
The optical setup is based on a Nikon  Eclipse Inverted Microscope. 
The channel is placed on a motorised stage, making it possible to track the 
rods over long time periods as they follow the  flow in the channel. 
The rods are observed using a 20X microscope objective (NA=0.28). Images are recorded with a CCD camera (Leica DFC 350FX, Switzerland) 
at a frame rate of 100 frames per second.
Each frame has a resolution of $348\times260$ pixels. The size of a pixel is $\unit[1.2]{\upmu m}\times\unit[1.2]{\upmu m}$.
As shown in Fig.~\ref{fig:experimental_setup}, the rods are observed through a cover glass. 
Since the magnification of the microscope is only 20X, the thickness of the glass does not cause focal problems.
The channel is illuminated through the transparent PDMS.

\begin{figure}
\includegraphics[width=11.5cm]{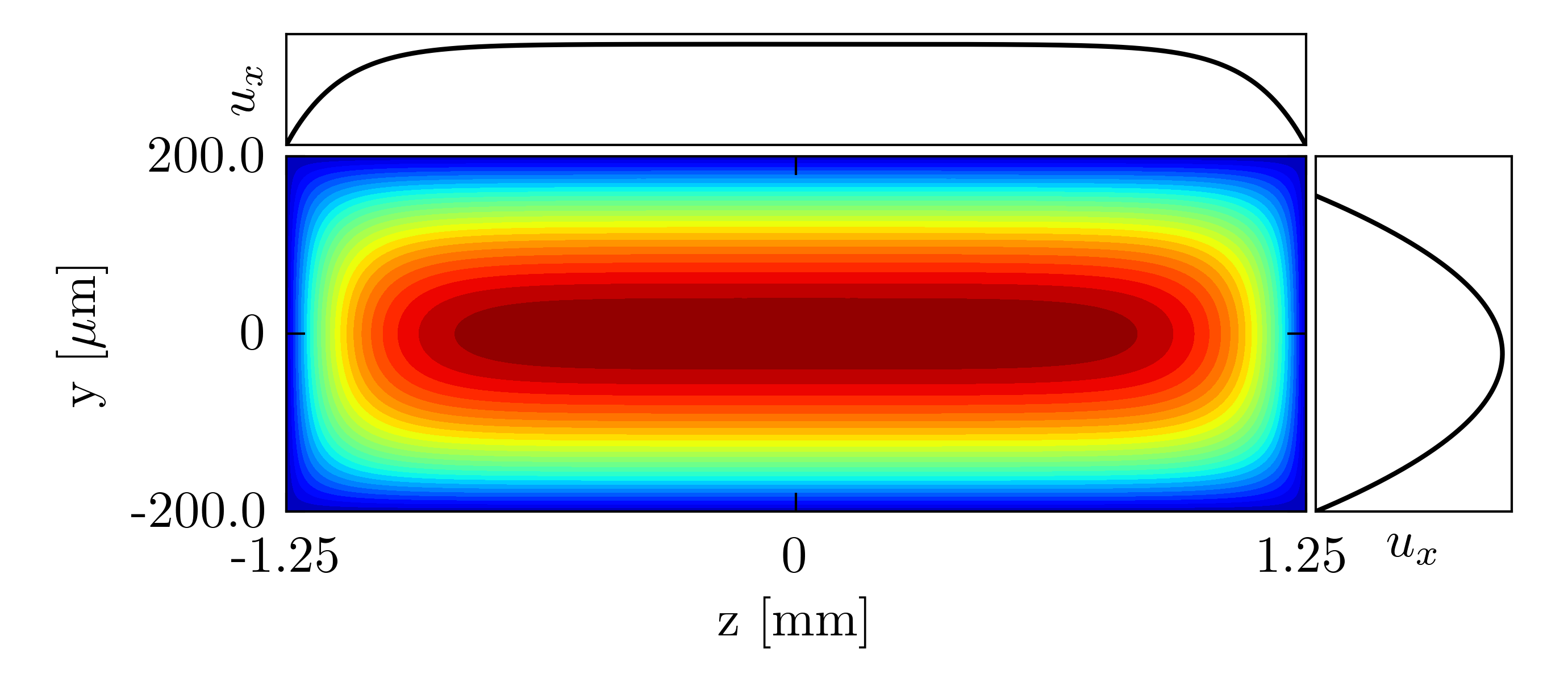}
\caption{\label{fig:flowprofile_theory} Computed flow profile for the channel geometry used in experiments. Width ($z$) of channel is $\unit[2.5]{mm}$ 
and depth ($y$) is $\unit[400]{\upmu m}$ (refer to Fig.~\ref{fig:experimental_setup} for coordinate system). The intensity plot shows a 
cross-section of the channel, intensity indicating flow velocity in arbitrary units. Top and right panels show the corresponding 
flow-velocity profiles at $y=0$ and at $z=0$, respectively. }
\end{figure}

Fig.~\ref{fig:experimental_setup} also shows the Cartesian coordinate system adopted in this paper. The $x$-axis is taken to lie along the flow in the channel.  The $y$-axis lies along the optical axis of the microscope (the channel is $\unit[400]{\upmu m}$  deep in this direction). The $z$-axis, finally, lies along the width of the channel (it is
\unit[2.5]{mm} wide). Thus the camera plane corresponds to the $x$-$z$-plane, and the channel cross section (Fig.~\ref{fig:flowprofile_theory}) corresponds to the $y$-$z$-plane.

{\em Channel flow}.  
The microchannel flow is pressure-driven, using a syringe pump.
The flow direction is periodically reverted. 
The flow is characterised by a very small Reynolds number. The rods are suspended in a 2:1 mixture of glycerol and water, 
corresponding to a kinematic viscosity of $\nu = \unit[2\cdot10^{-5}]{m^2/s}$ \citep{cheng2008}.
The flow speed is of the order of $\unit[100]{\upmu m/s}$.  Based on the smallest channel dimension 
($\unit[400]{\upmu m}$) this yields a Reynolds number of the order of Re = $10^{-3}$. 
The Navier-Stokes equations for the incompressible flow in the channel thus reduce to the viscous Stokes equation
which can be simply solved by a basis expansion. Closely following \citet{brody1996} we have computed
the flow profile in a cross section far from the in- and outlets (so that we can neglect the $x$-coordinate).
We assume no-slip boundary conditions at the channel walls and a constant pressure gradient over the channel length.
The resulting profile for our channel geometry is shown in Fig.~\ref{fig:flowprofile_theory}. 
The resulting flow velocity scales as $u_x \sim  \Delta p/\eta$ as a function of pressure difference $\Delta p$ 
and dynamic viscosity $\eta$.
But the form of the profile is not affected by the values of $\Delta p$ or $\eta$. In Fig.~\ref{fig:flowprofile_theory} we observe
that close to the centre of the channel, the flow gradient is oriented along the $y$-axis. 
In the following we refer to this axis as the \lq shear direction\rq. The $x$-axis is termed \lq flow direction\rq{}, and
the $z$-axis \lq vorticity direction\rq{}.

The rods analysed in this paper are between   $\unit[20]{\upmu m}$ to $\unit[30]{\upmu m}$ long.  
The viscous time scale $L^2/\nu$ is of the order of $10^{-5}$ s, which is much smaller than both the inverse shear rate and the flow-reversal time scale. The disturbance flow due to the the inclusion is therefore expected to obey the quasi-steady Stokes equation.

\subsection{Data analysis}
As explained below, the rods often tumble aperiodically. To classify different dynamical behaviours requires long time series. 
Since one traversal of the channel takes approximately $5$ minutes, large amounts of data must be analysed. We
have therefore designed  a data analysis software (in MATLAB) to track the centre-of-mass and orientational motion
of the suspended rods in an automated fashion. 

The raw data from the experiment consist of time-stamped positions of the motorised stage, together with frame sequences from the microscope. 
The analysis uses this information to estimate time series of the centre-of-mass position of a given rod, and the orientation of the unit vector $\vec{n}$ along its major axis:
\begin{equation}
\label{eq:n}
\vec{n} = (n_x,n_y,n_z)\,.
\end{equation}
The coordinate system is shown in Fig.~\ref{fig:experimental_setup}.

In the remainder of this section we briefly describe how the empirical data were processed.
First, we summarise  our algorithm for the image analysis. Second we explain how time is rescaled to
account for variations in the flow velocity due to reversals of the direction of the flow and due to fluctuations in the pump pressure.

\subsubsection{Image analysis}
The automated image-analysis software tracks the centre of mass and the orientation of a given rod 
as it moves in the channel. We separately discuss two steps: detecting the position and orientation of the rod,
and estimating its length.

{\em Finding the rod position and orientation}.  To begin with, static image noise is removed as follows.
For each pixel, the time-averaged intensity is computed, and for each time, the average intensity
of the frame is calculated. Then, for each pixel, the time-averaged intensity is subtracted from 
the instantaneous intensity, and the instantaneous frame average is added.

To locate the rod in a given frame, the software requires an initial guess of the centre-of-mass position.
The frame rate is sufficiently high so that the tracked rods move only slightly between subsequent frames. The initial guess is therefore taken to be the last-known rod position (for the first frame, manual user input is required).
\begin{figure}
\subfloat[\label{fig:data_analysisb}]{\includegraphics[height=2.5cm]{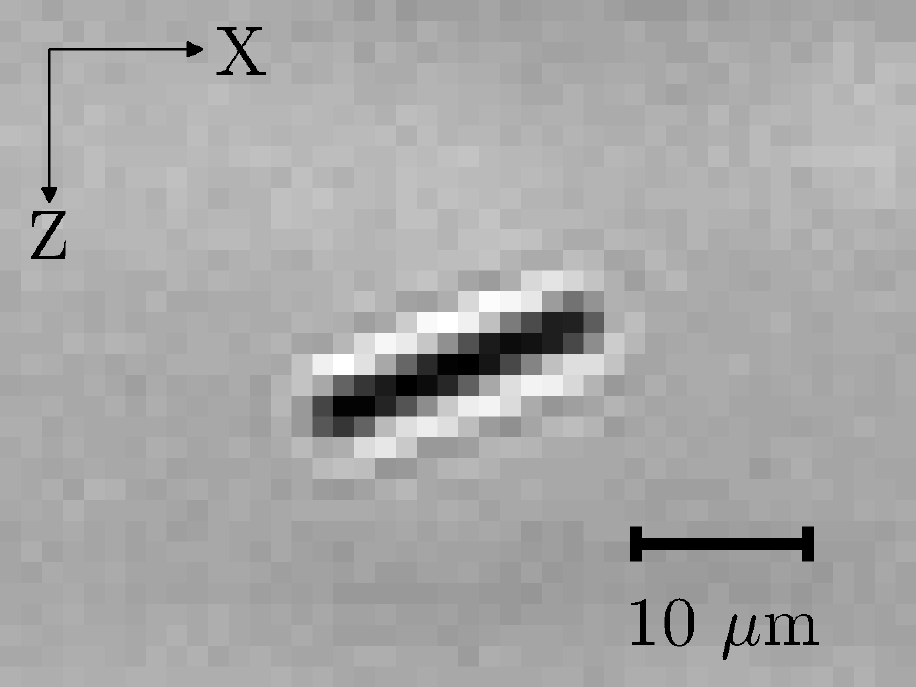}}\,
\subfloat[\label{fig:data_analysisc}]{\includegraphics[height=2.5cm]{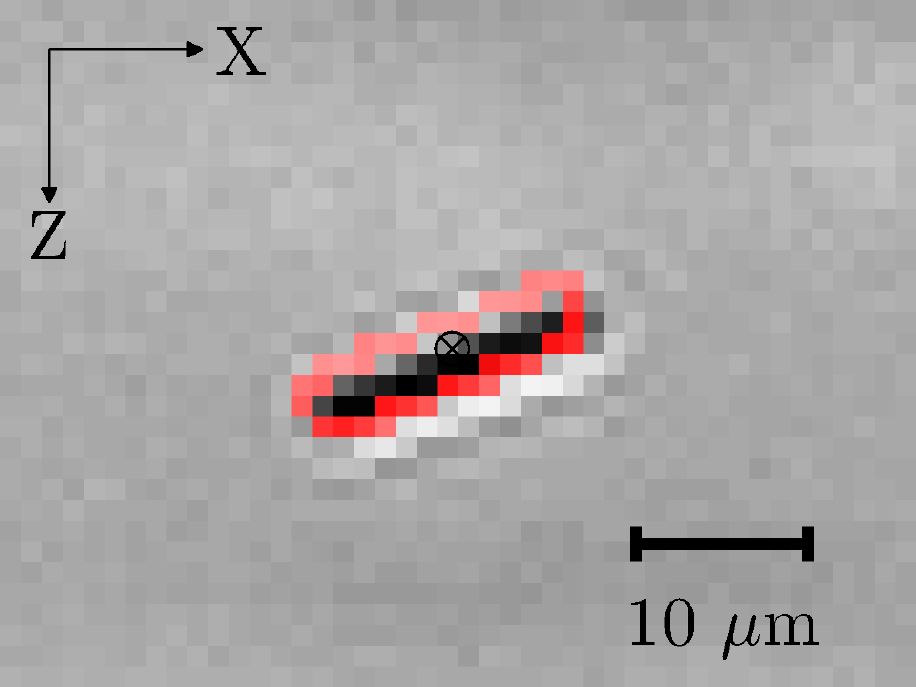}}\,
\subfloat[\label{fig:data_analysisd}]{\includegraphics[height=2.5cm]{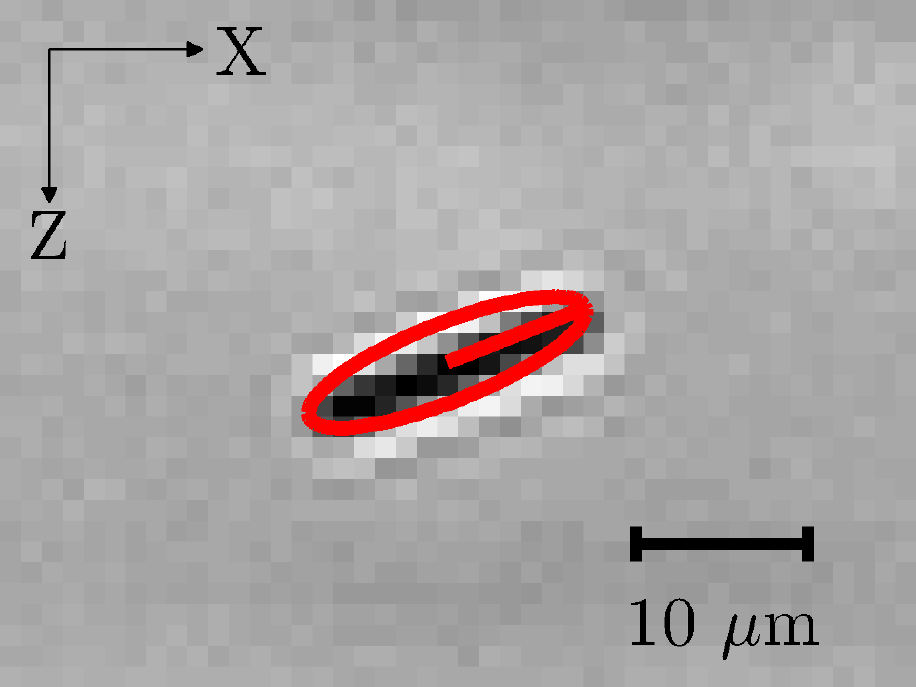}}
\caption{\label{fig:data_analysis} 
Images showing a small section of output from the  CCD,  cropped around the approximate position of the rod.      
The three images illustrate three steps in the data analysis: (a) the rod is located within the video frame, (b) edges are detected with 
an edge-detection algorithm \citep{canny1986}, (c) an ellipse is fitted to the resulting edge pixels. 
The orientation of the major axis of the ellipse is used to estimate 
orientation of the rod in the camera plane.}
\end{figure}

Once the location of the rod within a given frame has been estimated, a smaller window is cropped around 
the approximate position of the rod. An example is shown in Fig.~\ref{fig:data_analysisb}. 
A standard implementation of the edge-detection algorithm suggested by \citet{canny1986}  is used
to find the points on the edge of a rod sufficiently contrasted from its surroundings, as shown in Fig.~\ref{fig:data_analysisc}. 
We estimate the rod length and its orientation starting from the set of edge pixels. 
To this end we use least-squares fits to ellipses \citep{halir1998}, because this method has proven to be less sensitive to 
inaccuracies in the edge detection than the commonly employed principal component analysis. 
These fits yield estimates of the projections of ${\bf n}$ onto the camera plane (Fig.~\ref{fig:data_analysisd}).
Its $y$-component is not directly observable, but can be computed from $n_x$ and $n_z$ if the length
of the rod is known.

\begin{figure}
\includegraphics{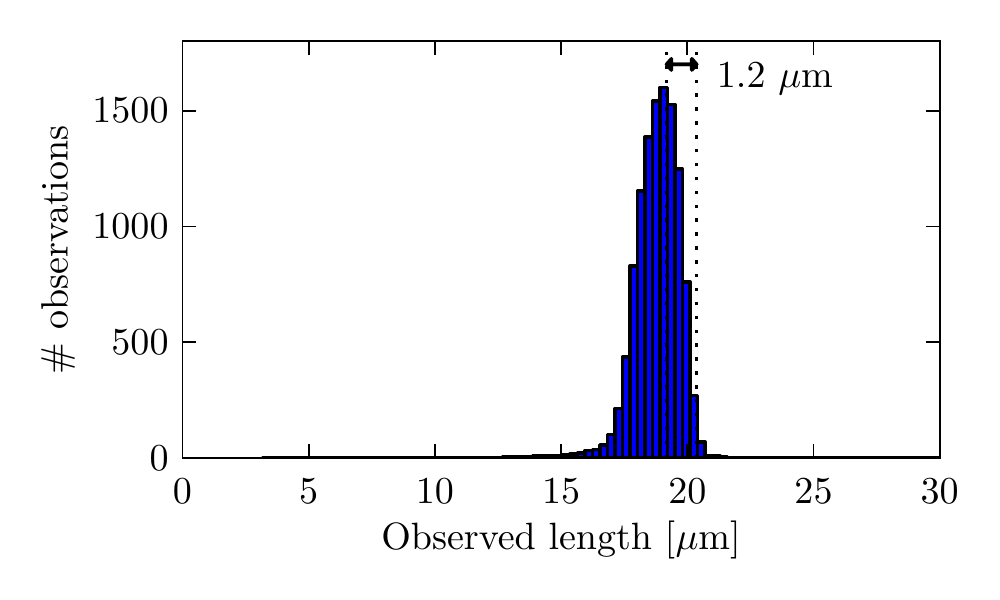}
\caption{\label{fig:length_histogram}%
A histogram of a series of measurements of the  projected rod lengths in the $x$-$z$-plane of a single rod (Rod \#1 in Section~\ref{sec:results}). The rod spends a long time nearly aligned with the flow direction. This results in the peak of the histogram. The tail to the left corresponds to observations when the rod is turning around, producing a shorter projection in the camera plane. The small tail to the right is due to the uncertainty in the  measurement. The width of the tail corresponds to the size a single pixel, indicated in the figure. The peak  position is taken to be the length of the rod.}
\end{figure}

{\em Estimating the rod length}.
It is impossible to determine the length of the rod from the length of its projected image in a single frame. 
However, the rod spends a significant amount of time aligned with the flow in the $x$-$z$-plane, where
the projected length corresponds to the rod length. We therefore recorded time series of projected rod lengths 
and computed the distribution of projected rod lengths (Fig.~\ref{fig:length_histogram}).
Ideally, the distribution should exhibit a sharp cut-off on its rhs, at the true length of the rod. 
However, we observe a small tail on the rhs of the distribution, caused by the experimental uncertainty of the 
order of one pixel size, or $\unit[1.2]{\upmu m}$. This uncertainty is indicated in Fig.~\ref{fig:length_histogram}. 
We therefore estimate the rod length by the position of the maximum of the distribution.

\subsubsection{Time rescaling}\label{sec:dimensionless}
As explained above, we revert the flow direction in order to test the sensitivity of observed orientational
motion to perturbations. We have found that small perturbations may have significant effects. It is therefore
important to revert the flow as smoothly as possible. This means that the flow velocity changes substantially 
over a significant amount of time. 
Another source of changes in flow speed $u_x$
is noise in the form of uncontrolled pressure fluctuations in the pump.

But as long as the flow is governed by the linear Stokes equation (as it is in our case), 
the only effect of fluctuations of $u_x$ is a linear change of time scale. To account for this change,
we plot orientational trajectories as a function of distance $d$ along the trajectory of the rod (and not
as a function of time $t$):
\begin{align}\label{eq:distance}
	d(t) &= \int_0^t \!\!\rd t'\,u_x(t')\,.
\end{align}
The instantaneous flow velocity $u_x$ is estimated by the centre-of-mass velocity. We assume, in other words,
that the centre-of-mass of the rod is advected by the channel flow. 
The change of variables (\ref{eq:distance}) greatly simplifies the analysis of our results. 

\section{Results}\label{sec:results}

\begin{figure}
\includegraphics{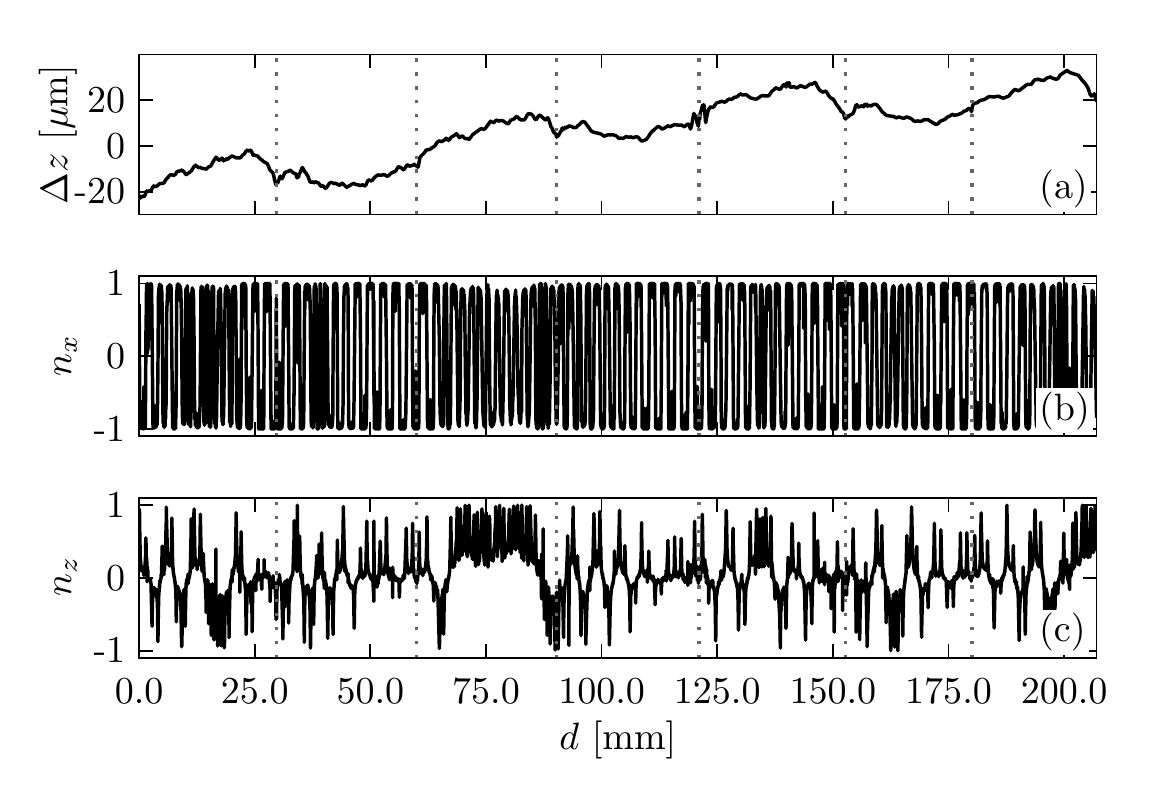}
 \caption{\label{fig:result_manystretches} Experimental output from the automated tracking system. Reversals of the flow direction are indicated by vertical dotted lines. (a) Relative centre-of-mass position in the $x$-$z$-plane of a rod advected in the microchannel. (b) 
$x$-component of orientation vector. (c) $z$-component of orientation vector.}
\end{figure}

Fig.~\ref{fig:result_manystretches} shows the output from the automated tracking system. The data show a long recording where the flow direction has been reversed several times by reverting the applied pressure difference. Each reversal is indicated by a vertical dotted line. The figure shows in panel (a) the relative displacement in the $z$-direction, and panels (b) and (c) show the $x$- and
$z$-components of the orientation vector, all as functions of distance the centre-of-mass is advected in the channel.

For a more detailed analys of the orientational dynamics we choose to show shorter trajectories. 
To measure the sensitivity of the dynamics to perturbations, we 
first track a given rod in one direction and record its centre-of-mass position and orientation.
Then we reverse the flow direction smoothly. In
Stokes' equation this turns $u_x$ to $-u_x$ and, by Eq.~\eqref{eq:distance}, $d$ to $-d$, effectively reversing time. 
In an ideal and noise-free experiment the centre-of-mass and orientation are expected to retrace their trajectories.
In reality, of course, and as clearly seen in Fig.~\ref{fig:result_manystretches}, we observe deviations 
that allow us to quantify the sensitivity of the observed dynamics to perturbations.

\begin{figure}
\includegraphics{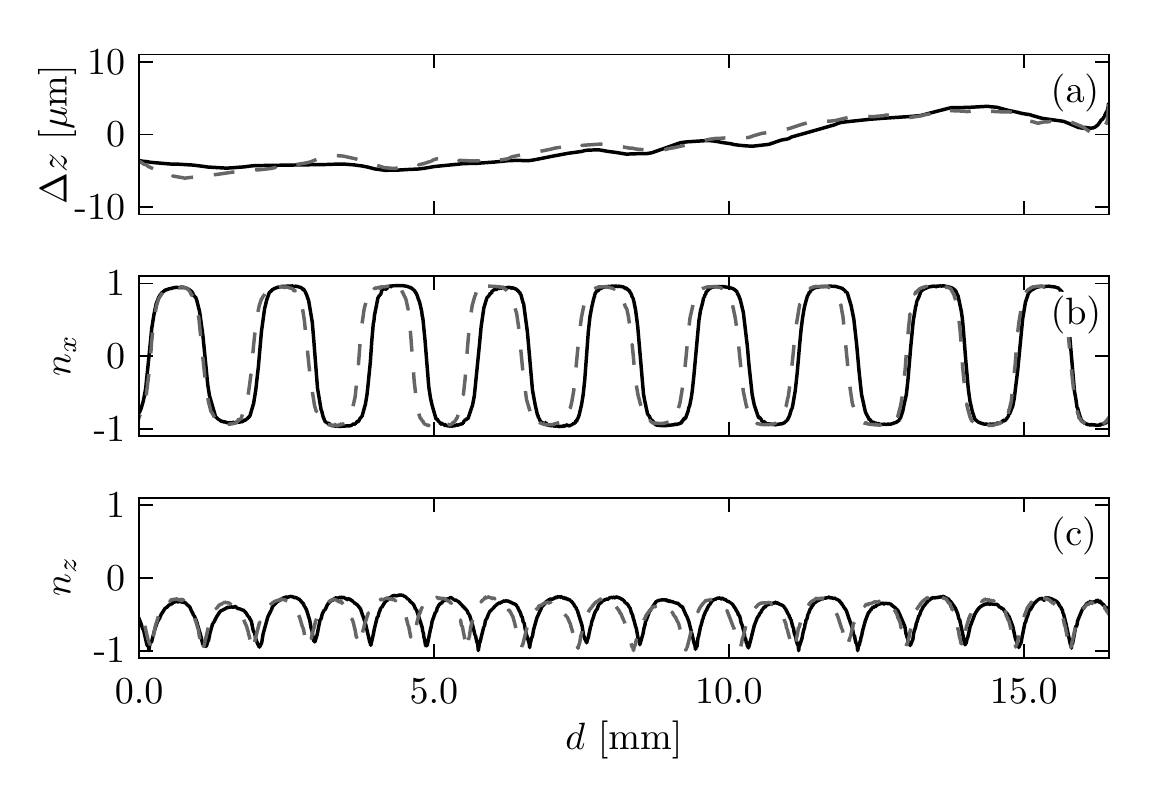}
 \caption{\label{fig:result_successful3} (a) Fluctuations $\Delta z$ 
of the $z$-coordinate of the centre-of-mass position of a rod advected in the microchannel flow from left to right (solid black line), reversed flow (dashed black line).
Here $d$ denotes the distance covered in the $x$-direction, see Eq.~(\ref{eq:distance}).
 (b) $x$-component of orientation vector. (c) $z$-component of orientation vector. 
 Rod \#1 (length  \unit[19]{$\upmu$m}, see Fig.~\ref{fig:length_histogram}).
}
\end{figure}

\begin{figure}
\includegraphics{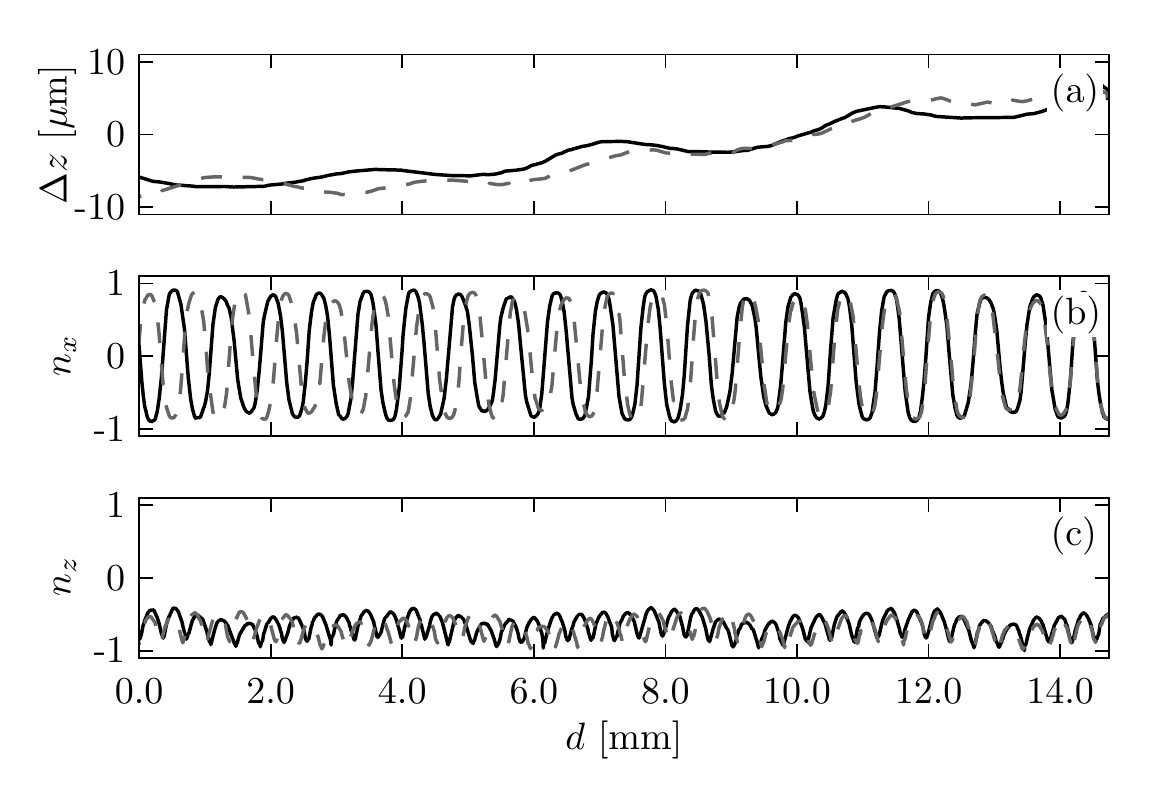}
 \caption{\label{fig:result_successful2} Same as Fig.~\ref{fig:result_successful3} but for rod \#2 (length  \unit[22]{$\upmu$m}).}
\end{figure}

\begin{figure}
\includegraphics{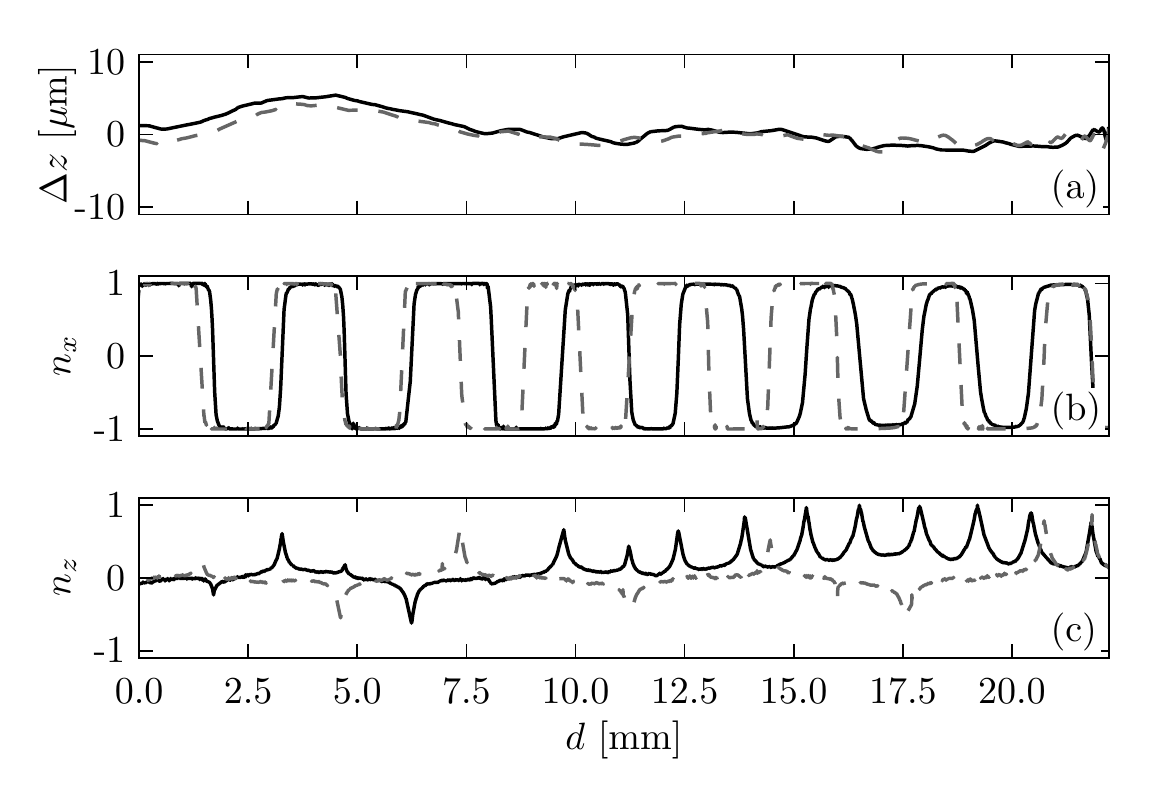}
\caption{\label{fig:result_butterfly1} 
Same as Fig.~\protect\ref{fig:result_successful3} but for rod \#3 (length  \unit[30]{$\upmu$m}).}
\end{figure}

Our results are summarised in Figs.~\ref{fig:result_successful3} to \ref{fig:result_butterfly1}. The data shown in 
the three figures were obtained for three different rods. Each figure contains three panels: (a) a high-resolution plot of the centre-of-mass trajectory
exhibiting small fluctuations $\Delta z$ in the $z$-direction, (b) the $x$-component of the orientation vector, 
before reversal (solid line), and after reversal (dashed line), and (c) the $z$-component of the orientation vector.

Consider first Fig.~\ref{fig:result_successful3}.
Panel (a) shows that the 
fluctuations $\Delta z$ of the centre-of-mass trajectory 
are very small, only on the order of $\unit[10]{\upmu m}$.
This is less than one rod length, and is much smaller than the channel width ($\unit[2500]{\upmu m})$. 
Panel (b) shows $n_x$, the component of orientation vector along the flow direction. 
The rod spends most of its time aligned with the flow, but regularly tumbles between $n_x=1$ and $n_x=-1$. Upon reversal, 
the orientational trajectory is initially similar, but drifts out of phase after several flips. The $z$-component of the orientation vector in panel (c) is, of course, close to $n_z=0$ while the rod is aligned along the flow. 
However, during a flip the values of $n_z$ determine which particular orientational path the rod takes.
We see that the shape of the $n_z$-curve during flips is similar for several subsequent flips. This indicates
that the orientational trajectory is approximately periodic.

Fig.~\ref{fig:result_successful2} shows a second example. 
It is similar to Fig.~\ref{fig:result_successful3}, but corresponds to a different orientational trajectory. 
The other main difference is that the period time of the tumbling motion is shorter.

Fig.~\ref{fig:result_butterfly1} shows an orientational trajectory
that exhibits large deviations after reverting the flow. 
The orientational motion is at first retraced, but after the rod has retraced
a full flip, deviations in the orientational motion
become noticeable and continue to grow rapidly. 

\section{Discussion}\label{sec:discussion}

Figs.~\ref{fig:result_successful3} to \ref{fig:result_butterfly1}
show both translational and orientational trajectories of rods advected in a channel flow. The translational trajectories indicate that the experiment is successfully controlled: the rods stay on straight streamlines during the whole experiment in Figs.~\ref{fig:result_successful3} to \ref{fig:result_butterfly1}. On a finer scale we see that the streamline followed by the centre-of-mass is slightly curved, on the order of $\unit[10]{\upmu m}$. This is probably
due to imperfections in the channel walls. After reversal, 
the centre-of-mass trajectories of the rods follow the streamlines very well, with only minor fluctuations on the order of one micrometer.

Our main results are the orientational trajectories of the rods shown in
Figs.~\ref{fig:result_successful3} to \ref{fig:result_butterfly1}.
As expected we observe tumbling motion, and the results include both periodic tumbling (Figs.~\ref{fig:result_successful3},\ref{fig:result_successful2}) 
and aperiodic tumbling (Fig.~\ref{fig:result_butterfly1}). Different realisations of the experiment result in different tumbling periods. 
Moreover, we observe a weak phase drift between trajectories before and after reversal.

We discuss these observations by first briefly recalling 
the hydrodynamic theory of \cite{jeffery1922}  and its implications for both axisymmetric and triaxial rods. 
The theory is valid when the difference in flow velocity at different ends of the rod is well approximated by the flow gradient. In our experiment this is the case because the length of the rods ($\unit[30]{\upmu m}$) is much smaller than the smallest channel dimension of $\unit[400]{\upmu m}$ (see also Fig.~\ref{fig:flowprofile_theory}).

Small axisymmetric rods suspended in a shear flow tumble periodically. 
More precisely, the orientation vector  
follows one of infinitely many possible periodic Jeffery orbits. 
Which particular orbit the rod takes is determined by its initial orientation.  
For most Jeffery orbits, the rod tumbles: it stays 
aligned with the streamline for most of the time, but at times
it rapidly flips by 180 degrees. 
When the initial orientation
of the rod is very close to the vorticity direction, 
the tumbling is less pronounced. The corresponding type of orientational 
motion is often referred to as ``kayaking'' in the literature, 
owing to its similarity to the motion of a kayak paddle.

For Jeffery orbits, the tumbling period does not depend on the initial orientation of the rod.
The period for a rod of aspect ratio $\lambda$ 
suspended in a shear flow of strength $s$ is given by
\begin{equation}\label{eq:jeffery_time}
	T_{\mathrm{Jeffery}} = \pi \frac{\lambda^2 + 1}{s \lambda}\,.
\end{equation}
As mentioned in the introduction, asymmetric rods also tumble, 
but not necessarily in a periodic fashion \citep{hinch1979, yarin1997}.

We begin the discussion of 
Figs.~\ref{fig:result_successful3} and \ref{fig:result_successful2}
by noting two consequences of Eq.~(\ref{eq:jeffery_time}).
First, this equation shows that the different tumbling periods
in Figs.~\ref{fig:result_successful3} and \ref{fig:result_successful2}
can result from differences in aspect ratio or shear strength:
larger aspect ratios correspond to longer periods, stronger
shear to shorter periods. The lengths of the
rods shown in 
Figs.~\ref{fig:result_successful3} and \ref{fig:result_successful2} are approximately the same. But the resolution 
of our microscope does not
allow us to precisely determine the thickness of the two rods.
We cannot exclude that the different tumbling periods are
in part caused by different aspect ratios of the two rods.

Second, we discuss the phase drift between the forward
and backward orientational trajectories seen in Figs.~\ref{fig:result_successful3} and \ref{fig:result_successful2} in terms of Eq.~(\ref{eq:jeffery_time}). 
We argue that the phase drift is due to fluctuations in the shear that are caused by 
small fluctuations of the $y$-coordinate of the centre-of-mass position.
As explained in Sec.~\ref{sec:experimental_methods}, the shear is a function of the $y$-coordinate. 
Our experimental set up does not allow us to measure fluctuations in the $y$-coordinate, but if
 we assume that the centre-of-mass drift in $y$ is similar in magnitude to that in $z$ (on the order of a few micrometers),
we can find a lower bound for the expected phase drift as follows. According to Eq.~\eqref{eq:jeffery_time}, the period is inversely proportional to the shear strength. 
It follows that the relative change in shear strength equals the relative change in time between flips, $\Delta s/s = \Delta T/T$. 
The flow profile in the $y$-direction is approximately a quadratic function, $u_x = -\alpha y^2$, which implies that $\Delta T/T = \Delta y/y$. 
Equating $y$ with the channel depth ($\unit[400]{\upmu m}$) allows us to estimate that the drift should at least be $\unit[1]{\%}$, 
in agreement with panels (b) and (c) in Figs.~\ref{fig:result_successful3} and \ref{fig:result_successful2}.

We conclude the discussion by examining the trajectory in Fig.~\ref{fig:result_butterfly1}. We see, first, that
the trajectory is clearly aperiodic, both before and after reversal. 
The numerical experiments by \citet{yarin1997} indicate that triaxial rods may exhibit aperiodic dynamics even for very small deviations from axisymmetry (that could
not be resolved by our microscope). Secondly, the forward and backward orientational trajectories in panel (c) of Fig.~\ref{fig:result_butterfly1} 
separate rapidly.  In fact, the trajectory appears at first to reverse perfectly but then the difference between forward and
backward orientational trajectories grows substantially. 
These observations suggest that Fig.~\ref{fig:result_butterfly1} corresponds to chaotic tumbling of a triaxial particle.

\section{Conclusions}\label{sec:conclusions}

We have designed a microfluidic setup with video microscopy and tracking software to measure the  translational and orientational trajectories of 
microrods advected in microchannel flows. The experiments presented here demonstrate the level of control 
and accuracy of the current setup. We observe both periodic and aperiodic (and possibly chaotic)
orientational dynamics.

Further work on the experiment aims to improve the efficiency and capability of the current setup. For example,
we plan to install an optical tweezer in order to control initial orientations
of the rods. This would make it possible to observe different orientational
behaviours (such as those shown in Figs.~\ref{fig:result_successful3} and \ref{fig:result_butterfly1}) for {\em the same} rod.

\begin{acknowledgements}
Financial support from the Swedish Research Council, the G\"oran Gustafsson Foundation for Research in Natural Sciences and Medicine,
and the European COST action MP0806 is gratefully acknowledged.
\end{acknowledgements}


\begin{thebibliography}{19}
\expandafter\ifx\csname natexlab\endcsname\relax\def\natexlab#1{#1}\fi
\expandafter\ifx\csname url\endcsname\relax
  \def\url#1{\texttt{#1}}\fi
\expandafter\ifx\csname urlprefix\endcsname\relax\def\urlprefix{URL }\fi

\bibitem[{Alargova et~al.(2004)Alargova, Bhatt, Paunov, and
  Velev}]{alargova2004}
Alargova, R., Bhatt, K., Paunov, V., Velev, O., 2004. Scalable synthesis of a
  new class of polymer microrods by a liquid-liquid dispersion technique.
  Advanced Materials 16~(18), 1653--1657.

\bibitem[{Bezuglyy et~al.(2010)Bezuglyy, Mehlig, and Wilkinson}]{bezuglyy2010}
Bezuglyy, V., Mehlig, B., Wilkinson, M., 2010. Poincar\'e indices of rheoscopic
  visualisations. Europhysics Letters 89~(3), 34003.

\bibitem[{Brody et~al.(1996)Brody, Yager, Goldstein, and Austin}]{brody1996}
Brody, J., Yager, P., Goldstein, R., Austin, R., 1996. Biotechnology at low
  Reynolds numbers. Biophysical Journal 71~(6), 3430--3441.

\bibitem[{Canny(1986)}]{canny1986}
Canny, J., 1986. A computational approach to edge detection. Pattern
  Analysis and Machine Intelligence, IEEE Transactions on PAMI-8~(6), 679
  --698.

\bibitem[{Cheng(2008)}]{cheng2008}
Cheng, N.-S., 2008. Formula for the viscosity of a glycerol-water mixture.
  Industrial \& Engineering Chemistry Research 47~(9), 3285--3288.

\bibitem[{Goldsmith and Mason(1962)}]{goldsmith1962}
Goldsmith, H., Mason, S., 1962. The flow of suspensions through tubes. i.
  single spheres, rods, and discs. Journal of Colloid Science 17~(5), 448 --
  476.

\bibitem[{Hal\'i\v r and Flusser(1998)}]{halir1998}
Hal\'i\v r, R., Flusser, J., 1998. {Numerically stable direct least-squares
  fitting of ellipses}. In: Proceedings of the 6th International Conference in
  Central Europe on Computer Graphics and Visualization. pp. 125--132.

\bibitem[{Hinch and Leal(1972)}]{hinch1972}
Hinch, E.~J., Leal, L.~G., 1972. The effect of Brownian motion on the
  rheological properties of a suspension of non-spherical particles. Journal of
  Fluid Mechanics 52~(04), 683--712.

\bibitem[{Hinch and Leal(1979)}]{hinch1979}
Hinch, E.~J., Leal, L.~G., 1979. Rotation of small non-axisymmetric particles
  in a simple shear flow. Journal of Fluid Mechanics 92~(03), 591--607.

\bibitem[{Jeffery(1922)}]{jeffery1922}
Jeffery, G.~B., 1922. The motion of ellipsoidal particles immersed in a viscous
  fluid. Proceedings of the Royal Society of London. Series A 102~(715),
  161--179.

\bibitem[{Kaya and Koser(2009)}]{kaya2009}
Kaya, T., Koser, H., 2009. Characterization of hydrodynamic surface
  interactions of \textit{Escherichia coli} cell bodies in shear flow. Phys.
  Rev. Lett. 103, 138103.

\bibitem[{Mishra et~al.(2012)Mishra, Einarsson, John, Andersson, Mehlig, and
  Hanstorp}]{mishra2012}
Mishra, Y.~N., Einarsson, J., John, O.~A., Andersson, P., Mehlig, B., Hanstorp,
  D., 2012. A microfluidic device for the study of the orientational dynamics
  of microrods. Vol. 8251. SPIE, p. 825109.

\bibitem[{Ott(1993)}]{Ott}
Ott, E., 1993. Chaos in Dynamical Systems. Cambridge University Press.

\bibitem[{Parsa et~al.(2012)Parsa, Calzavarini, Toschi, and Voth}]{Parsa}
Parsa, S., Calzavarini, E., Toschi, F., Voth, G.~A., 2012. Rotation rate of
  rods in turbulent fluid flow. Phys. Rev. Lett. 109, 134601.

\bibitem[{Petrie(1999)}]{petrie1999}
Petrie, C.~J., 1999. The rheology of fibre suspensions. Journal of
  Non-Newtonian Fluid Mechanics 87~(2-3), 369 -- 402.

\bibitem[{Wilkinson et~al.(2009)Wilkinson, Bezuglyy, and
  Mehlig}]{wilkinson2009}
Wilkinson, M., Bezuglyy, V., Mehlig, B., 2009. Fingerprints of random flows?
  Physics of Fluids 21~(4), 043304.

\bibitem[{Wilkinson et~al.(2011)Wilkinson, Bezuglyy, and
  Mehlig}]{wilkinson2011}
Wilkinson, M., Bezuglyy, V., Mehlig, B., 2011. Emergent order in rheoscopic
  swirls. Journal of Fluid Mechanics 667, 158--187.

\bibitem[{Wilkinson and Kennard(2012)}]{Wilkinson12}
Wilkinson, M., Kennard, H.~R., 2012. A model for alignment between microscopic
  rods and vorticity. J. Phys. A: Math. Theor. 45, 455502.

\bibitem[{Yarin et~al.(1997)Yarin, Gottlieb, and Roisman}]{yarin1997}
Yarin, A., Gottlieb, O., Roisman, I., 1997. Chaotic rotation of triaxial
  ellipsoids in simple shear flow. Journal of Fluid Mechanics 340, 83--100.

\end{thebibliography}
\end{document}